\documentclass[journal]{IEEEtran}
\ifCLASSINFOpdf
  \usepackage[pdftex]{graphicx}
\else
\fi
\usepackage[cmex10]{amsmath}
\begin{document}
%
\title{Fourier Plane Image Combination by Feathering}
%
%
%


\author{W.~D.~Cotton  
\thanks{National Radio Astronomy Observatory, 520 Edgemont Rd.,
Charlottesville, VA, 22903 USA email: bcotton@nrao.edu}
}

\maketitle

\begin{abstract}
Astronomical objects frequently exhibit structure over a wide range of
scales whereas many telescopes, especially interferometer arrays, only
sample a limited range of spatial scales.
In order to properly image these objects, images from a set of
instruments covering the range of scales may be needed.
These images then must be combined in a manner to recover all
spatial scales.
This paper describes the feathering technique for image combination in
the Fourier transform plane.
Implementations in several packages are discussed and 
{
example combinations of single dish and interferometric observations of
both simulated and celestial radio emission are given.
}
\end{abstract}


\section{Introduction}
%
%
%
%

\IEEEPARstart{C}{elestial} images may contain structure on a wide
variety of size scales whereas a given radio astronomical instrument
may be sensitive to only a limited range of size scales.
A common solution to this problem is to image the desired object with
a number of different instruments, or configurations of a given
instrument, to recover the structure on the total range of size scales
needed. 
These range from single dishes for the largest scales to short
baseline interferometers to longer baseline interferometers.
Furthermore, each instrument and/or array configuration may have
artifacts which are best dealt with using that data-set alone.
This paper considers the combination of images derived from several
instruments or configurations by the ``feathering'' technique.
{The term ``feathering'' is likely derived from the similarity with
bird's feathers which are dense at the center and very light at the edge.}
In this technique, images are combined in the Fourier transform
(``uv'') domain by a weighted average of the transform of the various
input images in order to extract the most appropriate spatial
frequencies from each. 

\section{Resolution and Spatial Dynamic Range}
Diffraction limited astronomical instruments have a resolution that is
proportional to the diameter of the aperture measured in wavelengths
of the light being observed.
The range of larger spatial scales to which the instrument is
sensitive depends on the details of the instrument and in particular,
the distribution and fraction of its aperture which is filled.
Filled aperture instruments (AKA ``single dishes'') sample all spatial
frequencies up to those defined by the total aperture.
Cost and other practical constraints limit the maximum size, hence
resolution of single dishes.
On the other hand, interferometers generally sample only a fraction of
the spatial frequencies less than that defined by the maximum
baseline.
The largest scale size that can be imaged is defined by the shortest
spacing that is adequately sampled.
Interferometers can be constructed to an arbitrary size, hence,
resolution but practical constraints limit the fraction of the
aperture that can be filled thus the largest scale size.
Some interferometers such as the VLA and ALMA have a ``zoom''
capability in which the antennas can be reconfigured to produce a
range of resolutions and surface brightness sensitivities.
In addition, ALMA has an array (ACA) of smaller antennas 
arranged in a more compact configuration to measure even lower spatial
frequencies. 
For objects with a wide range of spatial scales, multiple instruments
or configurations may be needed.

Every instrument can produce artifacts or spurious features in its
images.
Techniques to reduce these may be deployed but these are generally
specific to a given instrument or configuration.
An example of this is a bright source far from the pointing center of
an interferometer. 
Even when not in the field of interest, such a source may produce
side-lobes in the field of view that are subject to bandwidth smearing
and details of the far antenna pattern. 
Bandwidth smearing is locally convolutional and the effects of the
offending source can largely be removed from the data from a single
interferometer or configuration by deconvolution including that source.
Artifacts from asymmetries in the antenna pattern are not
convolutional but may be greatly reduced by corrections based on
known antenna patterns, by ``peeling'' or ``differential gains''
\cite{Smirnov2011}. 

Strong, extended sources whose structure {is} not adequately sampled by
an interferometer will have an extended negative ``bowl'' surrounding
them in derived images.
This is because the visibility at the center of the uv plane is the
total intensity in the field of view.
Thus, dirty images made lacking samples near the center of the uv plane
(which are implicitly replaced by zeros) will have an average value of
zero causing strong positive regions to be surrounded by negative
regions. 
The purpose of deconvolution is to interpolate between measured
visibilities including those near the origin.
If the uv coverage is inadequate to allow the deconvolution to recover
all of the emission in the source, portions of the bowl will remain.

\section{Feathering}
An image may be characterized by the spatial frequencies which are
well represented in the image; these correspond to the regions in the
Fourier transform (``uv'' space) of the image which were well measured
by the instrument.
Interferometers generally cover a range of spatial frequencies which
is limited by the longest and shortest baselines.
Single dishes, in principle, measure all spatial frequencies up to
those corresponding {to} the instrument diameter\footnote{However,
  spatial filtering to remove atmospheric or instrumental variations
  may filter out some spatial frequencies}.
Ideally images with overlapping, well--sampled spatial frequencies can
be combined to derive an image reproducing all the spatial frequencies
in the initial images.

Feathering is the technique of combining images in the uv plane to
recover the spatial frequencies in the input images.
The images to be combined must be adjusted to a common astrometric and
photometric scale and, if FFTs are to be used, interpolated onto a
common image grid.
The combination of $m$ images $I_i$ in the Fourier transform (u-v)
domain is a weighted average described by the following:  
$$ C(u,v) = \sum_{i=1}^m W_i(u,v)FT(I_i)(u,v) $$
where $FT$ denotes the Fourier transform, $W_i$ is the weighting
function for image $I_i$.
If there is emission near the edge of the images, 
{tapering the images to zero}
or similar techniques may be needed to reduce the artifacts
(ringing) resulting from the Fourier transform.
The feathered image is then:
$$ I(x,y) = N\ FT^{-1}(C)$$
where $N$ is a normalization factor.
The resolution of the resultant image should be that of the highest
resolution input image.
While this is a generic technique and is widely implemented, the
details, i.e. derivation of $W_i$ and $N$ and constraints on the
images vary from implementation to implementation. 

\section{Constraints on images}
Images to be combined need to be astrometrically and photometrically
aligned.
All implementation described below require the images to be
astrometrically aligned but differ in whether the photometric
alignment is including in the feathering.
In a range of cases it is possibly to align the flux density scales by
comparing the Fourier transform values in the annulus of overlapping
spatial frequencies.

Image combination will work best when the images combined have
{well sampled} overlapping regions of the uv--plane.
In the case of non-overlap, structure represented by the portions of uv
space not sampled will not be well represented in the feathered image.
Clearly, combining single dish and VLBI images will not generally be
productive. 

{Using the amplitude ratios in annuli may not always be the best
method for adusting the flux density scales but it is well defined for
single interferometric pointings and is widely used.
An alternative is to use isolated, spatially small features as an
adjustment of the astrometric grids is generally also needed and such
features can be used for both.

If the interferometer sampling in the uv overlap is really dense then
adjustment using the annuli should be relatively robust; unfortunately
this is often not the case and the interferometric measurements are
less well sampled. 
Since the comparison is made using the interferometric image rather
than the visibility data, the resolution of extended emission will
generally bias the amplitudes in the interferometric annulus low.
A correction for this will incorrectly increase the relative power in the
higher spatial frequencies making smaller scale structures erroneously
brighter. 
The annular method of adjusting flux density scale is more
problematic in the case of large scale images needing interferometric
mosaics. 
}

The technique described above assumes that each input image has been
appropriately filtered to remove power at spatial frequencies beyond
the range sampled by the instrument used to derive the image.
For interferometric images this constraint is generally that the
emission CLEANed is restored with a CLEAN beam that accurately
represents the psf of the instrument.
Images derived from single dish measurements may not have had similar
filtering and may need to have spatial frequencies outside of the
telescope aperture filtered before combination.

Interferometric images of sources with extended emission generally
need extra care as emission poorly sampled in the uv--plane can lead to
image artifacts.
Multi--resolution CLEAN can help with reconstructing large-scale
emission.
In extreme cases, setting a short baseline limit 
{during the imaging}
can help suppress large waves from very poorly sampled structure.
Negative bowls around extended emission are common but should be
removed by feathering if the relevant spatial frequencies are obtained
from other images.
\subsection{Primary beam correction}
Interferometric images should be ``Primary beam corrected'' before
feathering such that features in the image are at the strength they
would have without off--center attenuation of the power pattern of
the interferometer elements
\subsection{Mosaics}
Mosaics are suitable candidates for feathering as long as the
combination into the mosaic removes the primary beam pattern.
\subsection{Weighting}
Not all images are created equal and it may be desirable to allow
different images to have different weights where their uv regions
overlap.
A simple relative weight is the inverse variance of the noise.
{This additional weight is multiplied times the feathering weight.
}
\subsection{Spectral cubes}
Feathering spectral line cubes is a straightforward combination of each
set of channel images.
Cubes must either have a common channelization or be interpolated onto
one. 

\section{Alternatives to Feathering}
There are several alternatives to combining data from different
instruments.
Their properties differ and the best technique depends on the data
involved.
For a broad range of cases, feathering produces acceptable results.
See \cite{Stanimirovic2002} for a discussion on a variety of techniques
\subsection{Combine visibilities}
If several interferometers are to be combined, the visibilities can be
combined into a single data-set which is then imaged.
Single disk images can be Fourier transformed to produce ``pseudo''
visibilities. 
However, the single dish psf (antenna pattern) must first be
deconvolved and the result reconvolved with the interferometer antenna
pattern. 
Since the aperture is filled, this can be done using a linear
deconvolution.
AIPS task IM2UV can convert an image into pseudo visibilities.

Special care must be taken when combining data from interferometer
arrays with different antenna sizes as the field of view differs.
The imaging software must properly correct for the different antenna
sizes if a wide field of view or mosaic is to the imaged.

Visibility data is generally ``weighted'', that is each visibility is
given a weight with which it is to be used in the image construction.
These weights may need to be adjusted in the visibility combination
process to assure that the data makes the appropriate contribution to
the derived image.
\subsection{Combine dirty images and deconvolve}
The formation of dirty images is linear so an interferometer image can
be linearly combined with other interferometer dirty images or a
single dish image.
The combined image can then be deconvolved; Maximum Entropy (MEM) is
frequently used for this.
Miriad task MOSMEM applies this technique.
\subsection{Use single dish image  as model in deconvolution}
A discussion of using the single dish image as an initial model using
Maximum Entropy deconvolution is described in \cite{Cornwell1988}

\section{Implementations}
\subsection{AIPS Implementation}
The AIPS \cite{AIPS} implementation is in task IMERG and will combine
two images.
These must be on a common grid and size but IMERG will adjust the flux
density scale using the values in a specified annulus of the uv plane.
Interior to the specified uv annulus, the lower resolution image is given
weight 1 and the higher resolution image a weight 0.
These are reversed outside the annulus and within the annulus an
exponential is used to give increasing weight to the higher resolution
image. 
Since the sum of the weights are always 1, the normalization factor
is 1.

\subsection{CASA Implementation}
The CASA \cite{CASA} implementation is in task feather and allows 
combining two images and consists of the following steps:
\begin{itemize}
\item Interpolate the low resolution image onto the grid of the high
resolution image.
\item Fourier transform both images to the uv--plane.
\item The weight of the low resolution image is the ratio of the beam
  area of the high resolution image to that of the low resolution
  image
\item The weight of the high resolution image is 1-w, where w is the Fourier
transform of low resolution CLEAN beam.
\item Back transform combined grid to image plane.
\item No normalization is documented.
\end{itemize}
\subsection{Miriad Implementation}
The Miriad \cite{Miriad} implementation is in task immerge with option 'feather'
which allows combining two images with Gaussian psfs.
The input images must be on a common grid and size and flux density
scale although the program will optionally adjust the lower
resolution flux density scale using overlapping spatial frequencies.
The weight of the low resolution image is 1 everywhere and that of the
high resolution image is the Fourier transform of the low resolution
CLEAN beam.
The resultant resolution is that of the higher resolution image.

\subsection{Obit Implementation}
The feathering implementation in Obit\cite{OBIT}
\footnote{http://www.cv.nrao.edu/$\sim$bcotton/Obit.html} follows that
described in \cite{Farhad2004} and is implemented in task Feather
which will combine up to 10 images.
Feather requires only that the astrometric parameters
in each image give the same positions for given features
and need not be on the same grid, projection or even equinox.

The photometric (flux density) scale must be the same for all input
images. 
If this cannot be achieved from the calibration of the data, an
adjustment can be derived prior to feathering from the average ratio
of amplitudes in the overlapping region of uv space.

Images are presumed to have an annular region in the uv--plane which is
well sampled and a weighting mask is constructed for each which tapers
to zero at lower spatial frequencies.
{In the case where there is not an overlap in the well sampled
regions of uv space, structures corresponding to spatial frequencies
not well sampled in any of the input images may be poorly represented
in the feathered image, or missing entirely.
}
The weighting mask of the lowest resolution image is not tapered to
its center.
Feather depends on the ``CLEAN'' beam size given in
each image descriptor accurately reflecting the resolution of the
image.
The various steps in feathering are described in the following.

\begin{enumerate}
\item  {\bf Re-sample images }\\
The first step is to re-sample images with resolutions less than the
maximum to the grid of the maximum resolution image with sufficient
zero padding on the outside to allow an efficient FFT.
The interpolation uses the Lagrangian technique in 2D to interpolate
the pixels in the lower resolution images at the locations of the
highest resolution image using a 5x5 pixel  kernel.
\item  {\bf Generate weighting masks }\\
For each resolution except the lowest, a real weighting mask is
generated with a Gaussian hole in the center representing the spatial
frequency range of the next lowest resolution.
A sampling mask representing the spatial frequencies of each image is
generated by:
\begin{enumerate}
\item Create image with the CLEAN beam at the center 
\item FFT
\item Take real part
\item normalize to 1 at (0,0) spatial frequency
\end{enumerate}
The weighting mask for each image $i$ is 
$$ weight\_mask_i\ =\ 1.0 - sampling\_mask_{i+1}$$
where $i+1$ indicates the next lowest resolution.
The weighting mask for the lowest resolution is 1.0 everywhere.
The weighting masks are multiplied by the weights assigned to the
input images.
\item  {\bf FFT }\\
Each image is FFTed to the uv plane
\item  {\bf Weight }\\
Multiply Fourier transform of image by its weighting mask.
\item  {\bf Accumulate }\\
Sum the Fourier transform of the images times weight.
\item  {\bf Inverse FFT }\\
Fourier transform back to the image plane.
\item  {\bf Normalize } \\
The normalization factor is determined by repeating the process but
replacing the image with its corresponding CLEAN beam.
The normalization factor is 
{1.0/flux-density(center pixel) of the feathered beam.}
\end{enumerate}

\section{ Examples}
\subsection{Simulation}
{  
To illustrate the power of this technique, noiseless synthetic data sets
were derived for the model source distribution shown in Figure
\ref{ModelFig}.
A simulated single dish image was derived by convolving the model
image with the resolution of the single dish.
To simulate an interferometric image, a sample full track, VLA--like
array uv coverage  was generated (Obit/UVSim) and the Fourier
transform of the model was evaluated at the locations in the data-set.
These data were then imaged with multi--resolution CLEAN (Obit/Imager).
The simulated single dish and interferometric images are shown in
Figure \ref{ObsFig}.
The very extended emission is only visible in the single dish
representation and the most compact emission only visible in the
interferometric version.
These two images were then feathered together with equal weights using
Obit/Feather giving the image shown in Figure \ref{ModFeatherFig}.
The interferometer image contains 7\% of the initial model flux
density as derived by an integral over the image whereas the single dish
image has 96\%. 
The combined, feathered image also contains 96\% of the initial model
flux density.
The missing 4\% is due to truncating the single dish image.

\begin{figure}
\centerline{
  \includegraphics[width=3.5in]{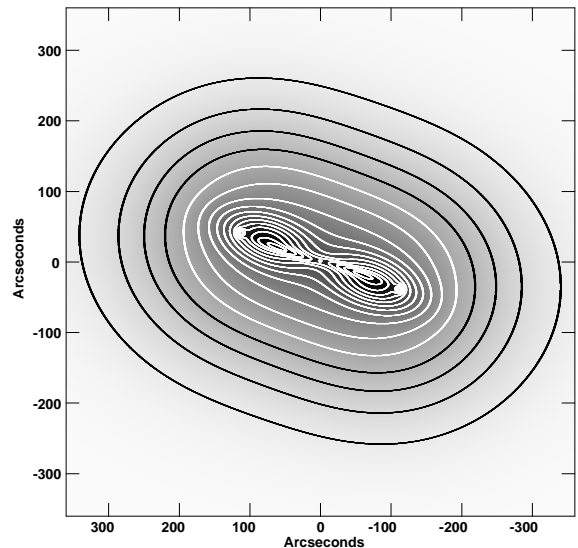}
}
\caption{Model source shown in reversed grayscale with linearly spaced
  contours overlaid.} 
\label{ModelFig}
\end{figure}
\begin{figure*}
\centerline{
  \includegraphics[width=3.5in]{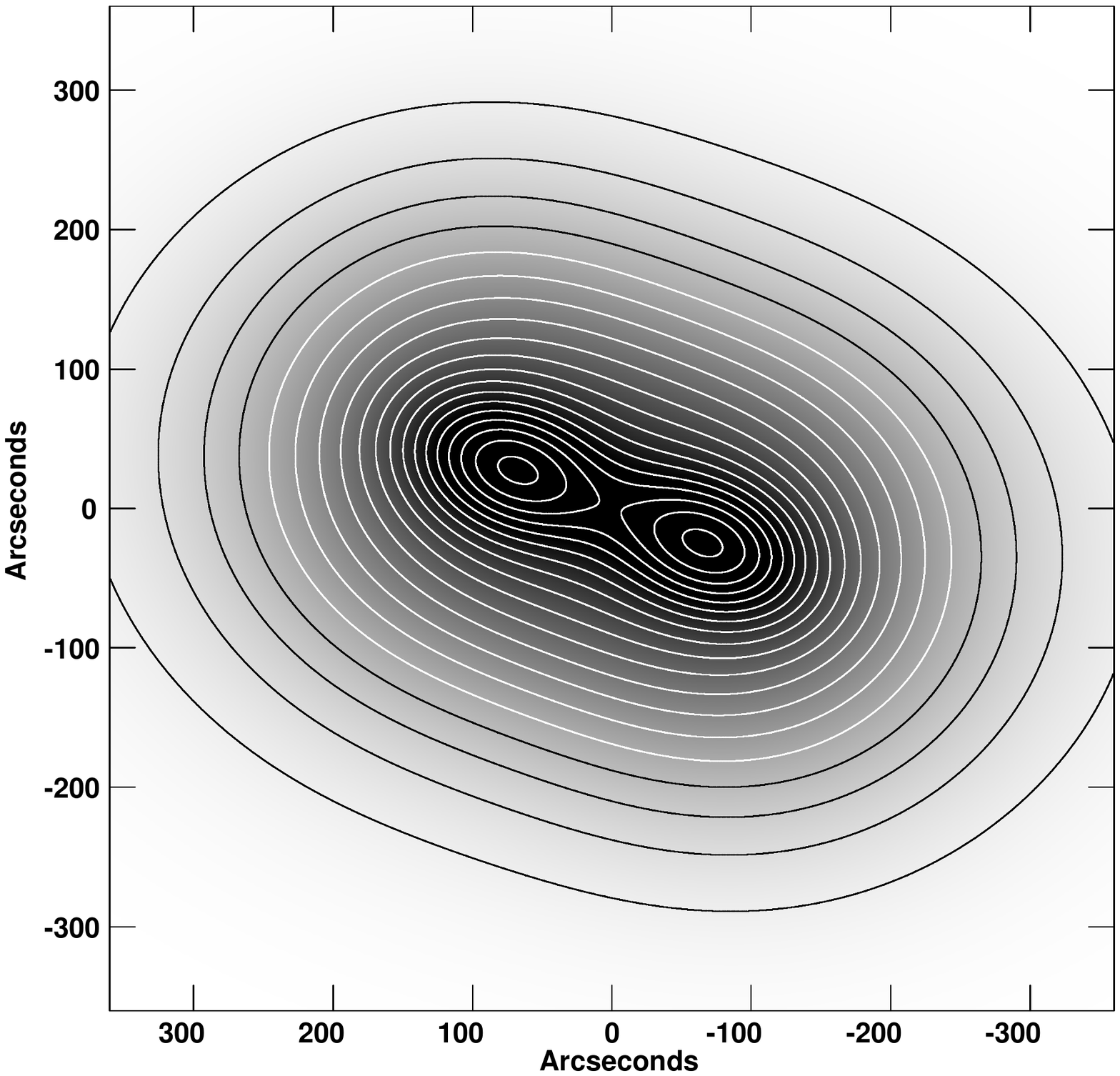}
  \includegraphics[width=3.5in]{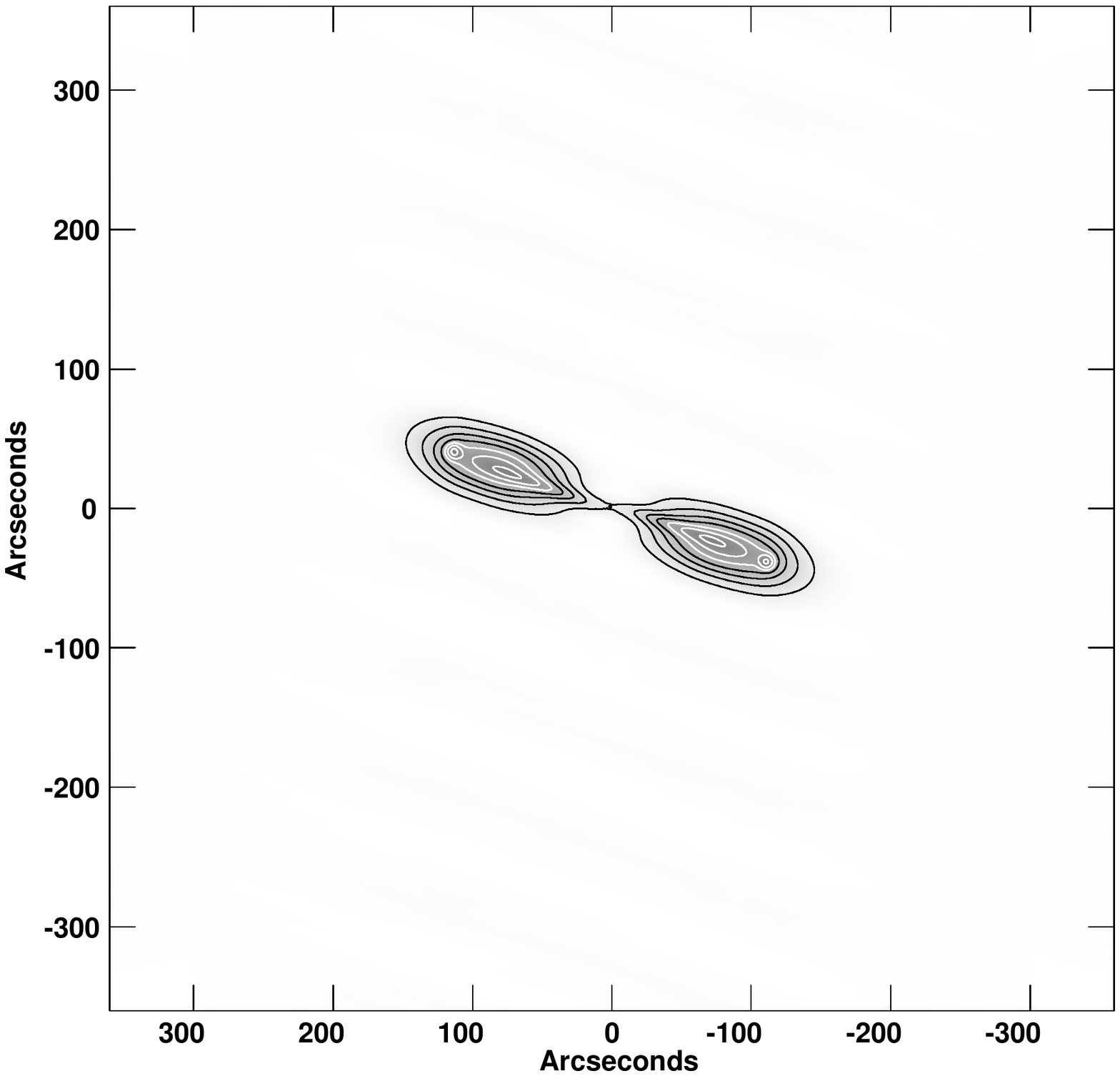}
}
\caption{On left is the model source as observed with a single dish
and on the right with an interferometer.
Contours are at the same levels as Figure \ref{ModelFig}.
} 
\label{ObsFig}
\end{figure*}
\begin{figure}
\centerline{
  \includegraphics[width=3.5in]{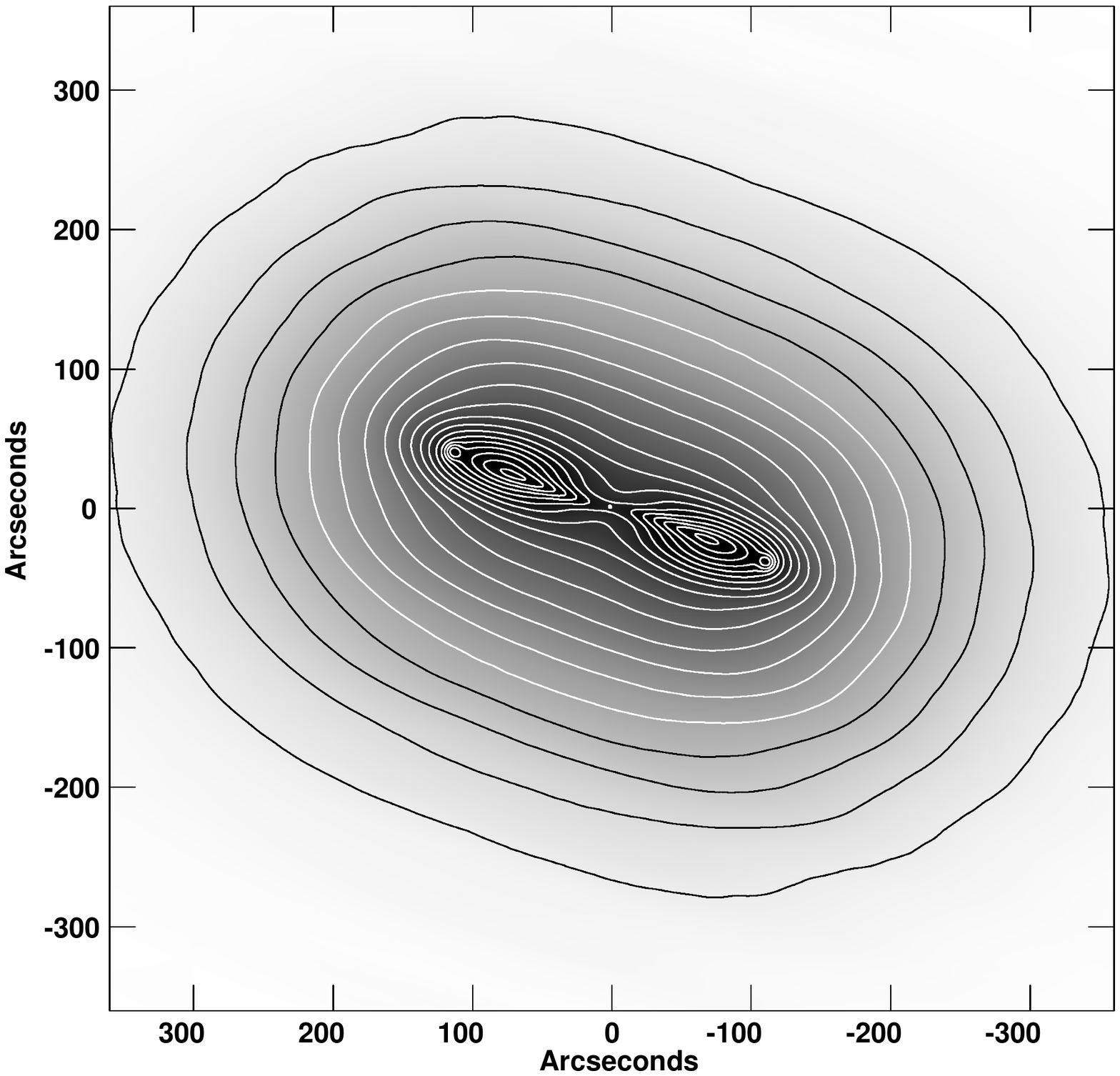}
}
\caption{The feathered combination of the images in Figure \ref{ObsFig}.
Contours are at the same levels as Figure \ref{ModelFig}.
} 
\label{ModFeatherFig}
\end{figure}
} 

\subsection{Celestial Example}
\begin{figure*}
\centerline{
  \includegraphics[width=3.5in]{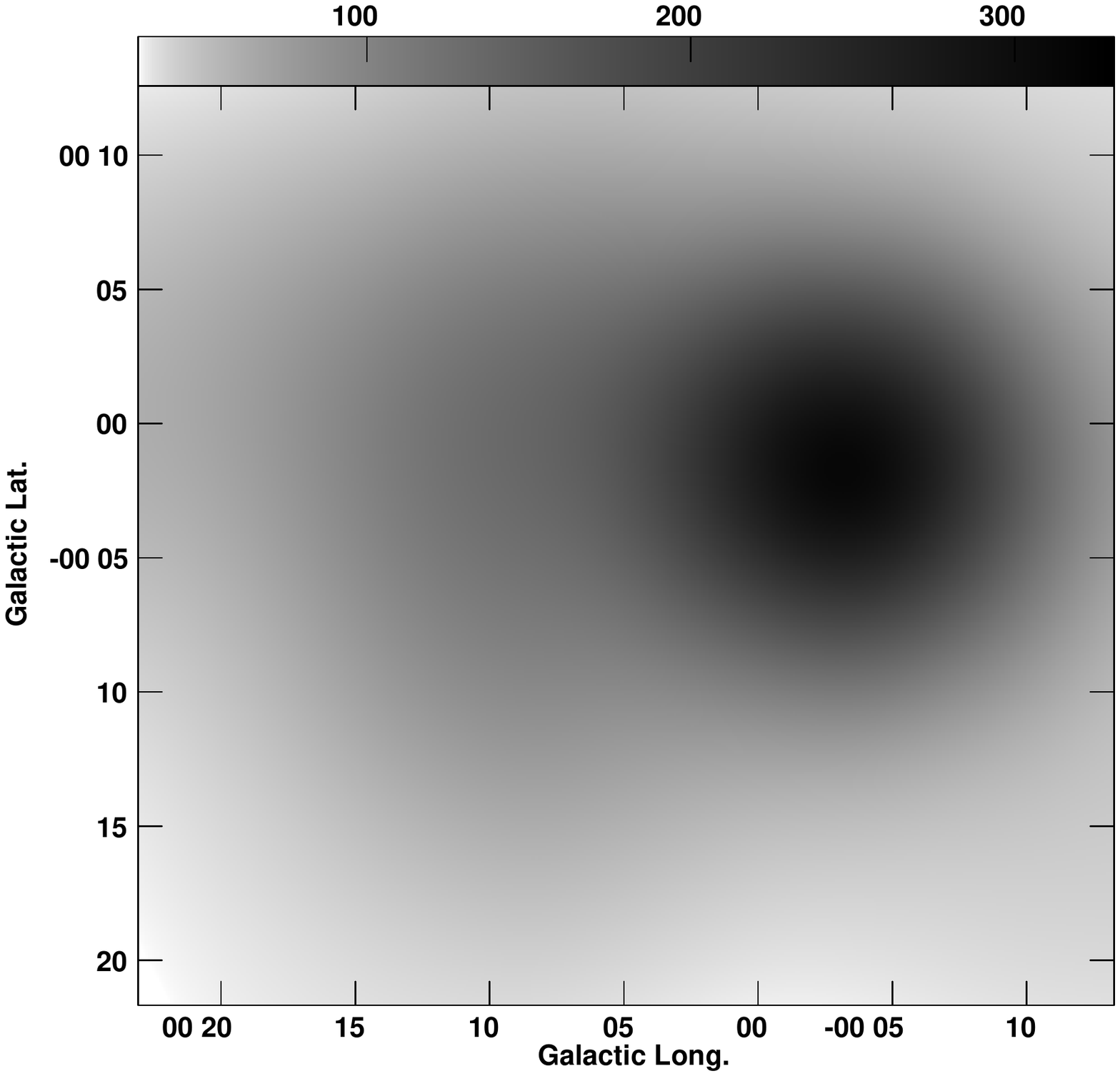}
  \includegraphics[width=3.5in]{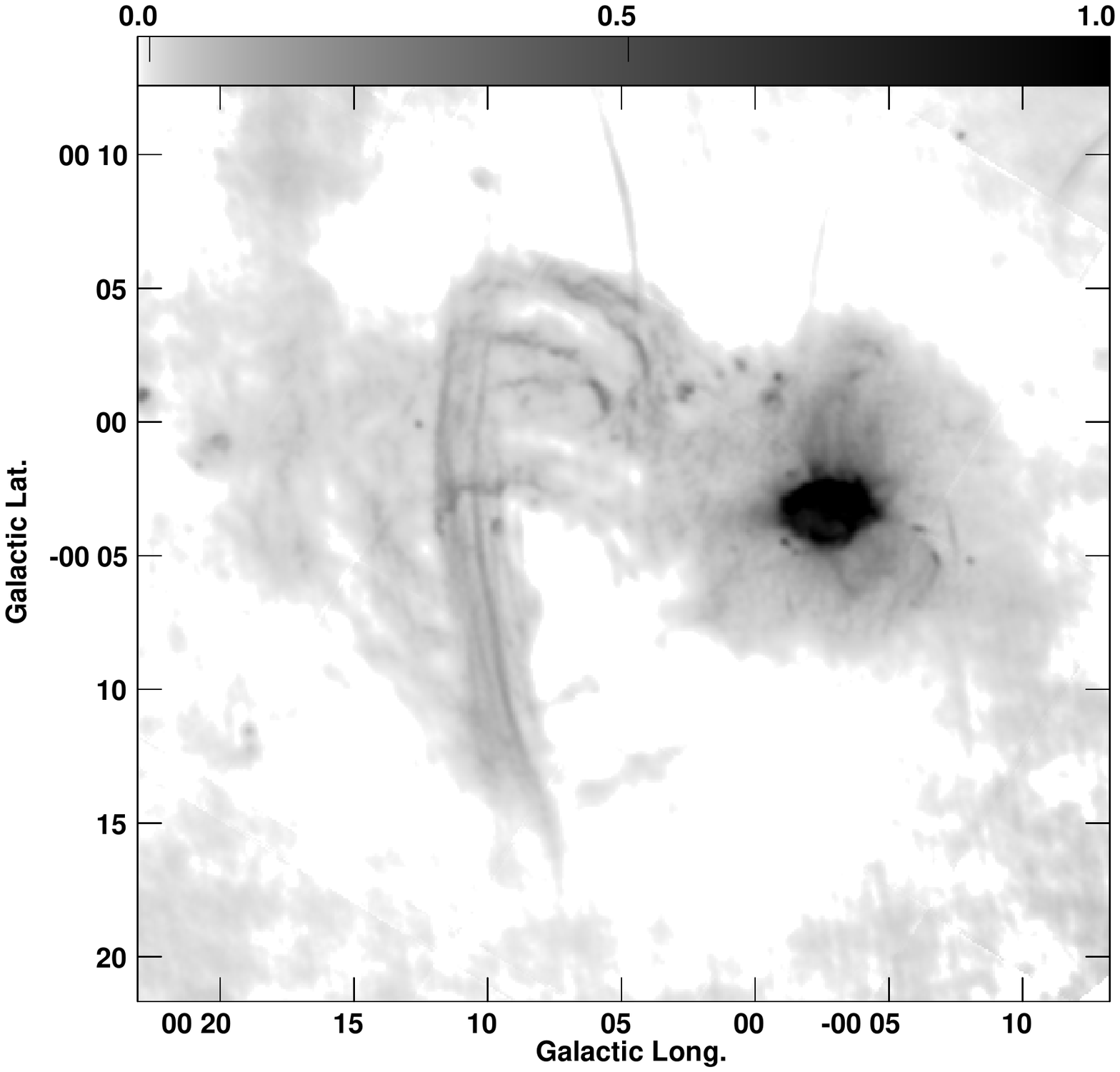}
}
\centerline{
  \includegraphics[width=3.5in]{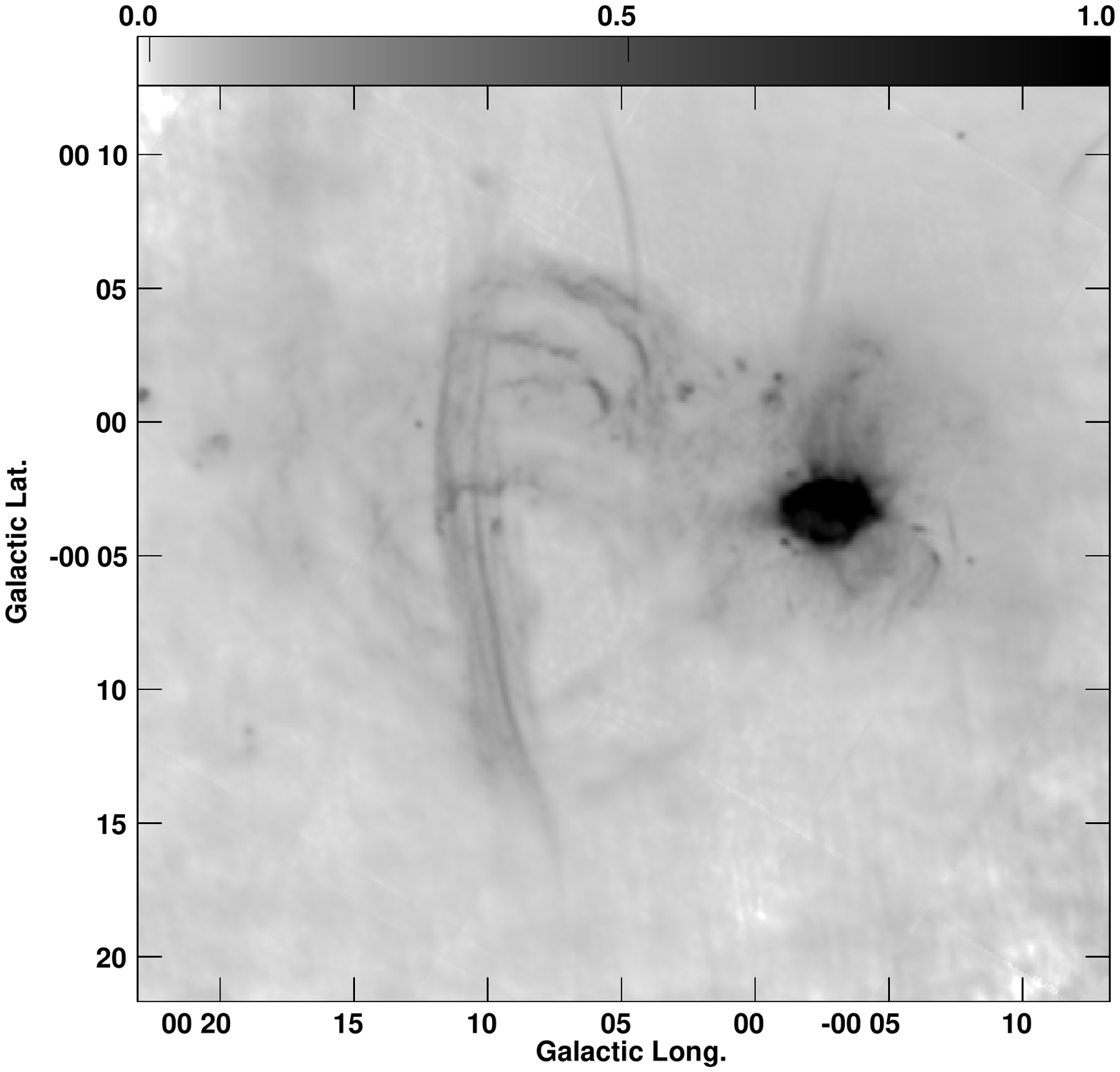}
}
\caption{Top: left is the GBT single dish image
and on the right the VLA interferometric image.
Grayscale stretch is square root with scale bar at
top.\hfil\break 
Bottom: The feathered combination of the images.
Grayscale stretch the same as top right.
Images are given as negative grayscale.
}
\label{FeatherFig}
\end{figure*}

For a celestial example, single dish observations of the Galactic
center at 20 cm wavelength made with the GBT \cite{Farhad2004} are
combined  with a VLA mosaic at the same wavelength \cite{LawGCA},
\cite{LawGCB}. 
Figure \ref{FeatherFig} shows a subset of the image around the
Galactic center with the single dish and interferometric images on
top and the combined image on the bottom.
The images were aligned astrometrically and photometrically using
several strong, isolated HII regions.
The missing short spacing in the VLA image leave a deep bowl around
the image which is filled in by adding the GBT image.
{
Due to the strong, wide spread emission in this part of the sky,
integrated flux densities cannot be accurately determined from this
data. 
Note: the VLA image is a mosaic derived in AIPS using a suboptimal
image plane ``feathering'' to combine the images resulting in the
lines visible between adjacent pointings.
These artifacts are greatly reduced if the spatially overlapping
images are feathered in the sense of weighting that decreases to zero
at the edge.
}

\section{Discussion}
The feathering technique for image combination in the Fourier plane 
is described and {examples} of its use shown. 
{In an example using simulated data, the feathered image recovered
all of the flux density in the low resolution image.}
In the {celestial} example a deep negative bowl around a bright
region of emission in the interferometric image is removed by adding
the single dish data. 

The feathering technique {as implemented} in the Obit package
described here allows an arbitrary number of images at different
scales to be combined. 
This allows the best approach to reducing artifacts to be used for
each of the multiple interferometric and single dish images before
their combination. 

\section*{Acknowledgment}
The author would like the thank the anonymous referee for comments
leading to an improved paper.
The National Radio Astronomy Observatory (NRAO) is operated by
Associated Universities Inc., under cooperative agreement with the
National Science  Foundation. 


\bibliographystyle{IEEEtran}
\bibliography{Feather}
%




\end{document}